\documentclass[aps,pre,showpacs]{revtex4}
\usepackage{graphicx}
\usepackage{dcolumn}
\usepackage{bm}
\usepackage{amsmath}

\begin{document}

\title{Wrinkling of a bilayer membrane}

\author{ A. Concha}
\author{J. W. McIver III}
\author{P. Mellado}
\author{D. Clarke}
\author{O. Tchernyshyov}
\author{R. L. Leheny}

\affiliation{Department of Physics and Astronomy, Johns Hopkins
University, Baltimore, Maryland 21218}

\date{\today}

\begin{abstract}

The buckling of elastic bodies is a common phenomenon in the
mechanics of solids.  Wrinkling of membranes can often be
interpreted as buckling under constraints that prohibit large
amplitude deformation.  We present a combination of analytic
calculations, experiments, and  simulations to understand wrinkling
patterns generated in a bilayer membrane.  The model membrane is
composed of a flexible spherical shell that is under tension and
that is circumscribed by a stiff, essentially incompressible strip
with bending modulus $B$.  When the tension is reduced sufficiently
to a value $\sigma$, the strip forms wrinkles with a uniform
wavelength found theoretically and experimentally to be $\lambda =
2\pi(B/\sigma)^{1/3}$.  Defects in this pattern appear for rapid
changes in tension.  Comparison between experiment and simulation
further shows that, with larger reduction of tension, a second
generation of wrinkles with longer wavelength appears only when $B$
is sufficiently small.

\end{abstract}


 \maketitle
 
 \section{Introduction}
 
Under excessive load a rigid body will buckle.   A classic analysis
of buckling dates to Euler who considered a thin rod under
compressive stress along its axis.  Above a critical stress the rod
deforms with a wavelength twice as long as the rod, as the system
relieves compressive energy at the expense of bending energy.
Variations of Euler buckling appear in a range of geometries;
however, under certain constraints, deviations from this behavior
occur, and shorter wavelength patterns emerge.  For example, rods
subjected to a sudden compression buckle dynamically with a
wavelength that scales inversely with the impacting
stress~\cite{gladden}.  A second example is the buckling of
microtubules in biological cells for which the wavelength is reduced
due to mechanical coupling to the surrounding cytoskeleton that
energetically penalizes large amplitude
deviations~\cite{brangwynne}.  A related, well studied phenomenon
occurs in the wrinkling of stiff elastic membranes attached to thick
soft substrates~\cite{allenbook,bowden98,jan2001,cerda02, mahawrinkle,
maha_natmat}.  Stretching of the substrate makes large amplitude, and 
hence long
wavelength, deformations prohibitive, and the membrane wrinkles with
a wavelength set by a compromise between the bending and stretching
energies.  A number of applications have been advanced that exploit
the resulting periodic modulation such as in diffraction
gratings~\cite{harrison}, stretchable
electronics~\cite{lacour,khang} and metrological
tools~\cite{metrology}, indicating the potential of membrane
wrinkling as an approach for patterning materials.

\begin{figure}
\includegraphics[scale=1.5] {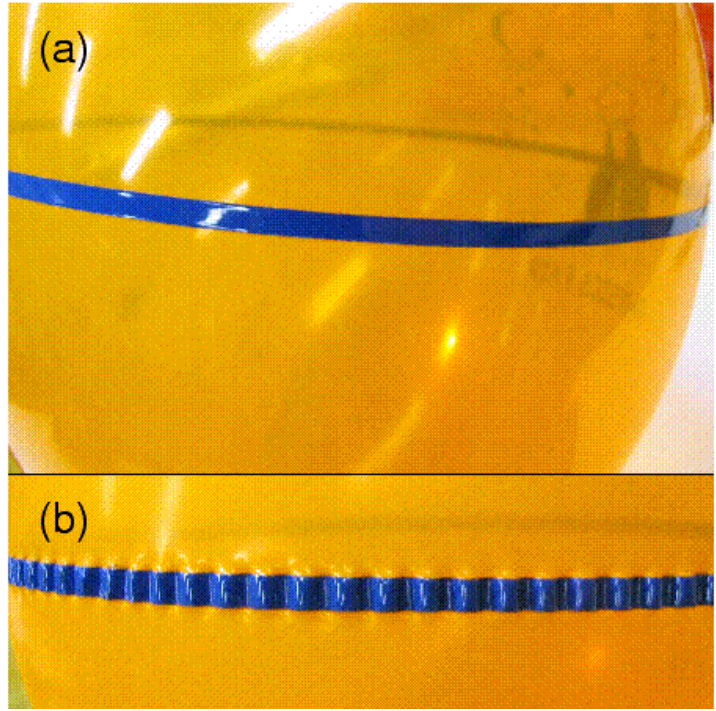}
\caption{\label{Figure 1:} (Color online) Experimental model of the bilayer
membrane, which is comprised of a stiff strip of paint
circumscribing an inflated rubber balloon.  (a) Initially, the strip
is free of stress and undeformed.  (b) When the balloon is partially
deflated, the strip wrinkles with a uniform wavelength. }
\label{fig1}
\end{figure}

In order to explore the variety of contexts in which one can controllably
wrinkle membranes, we present a model of a bilayer membrane with a
constraint that similarly prohibits long wavelength deformations.
The membrane is composed of an inner layer that is a flexible
spherical shell and an outer layer that is a stiff strip
circumscribing the sphere.   Initially, the sphere is stretched
under an isotropic tension, while the strip layer is free of stress.
When the tension on the sphere is reduced sufficiently, its radius
decreases, and the strip buckles with a well defined wavelength.  We
have investigated this bilayer system in simple experiments by
applying thin strips of paint around the equator of inflated rubber
balloons, as shown in the photograph in Fig.~1(a).  When the air
pressure in the balloon is reduced, the strip wrinkles with a
wavelength $\lambda$ that is much smaller than the radius of the
balloon, as illustrated in Fig.~1(b).  Comparing measurements of
this wavelength with quantitative analytic results, we determine the
dependence of $\lambda$ on other measurable parameters.  Notably, we
find that $\lambda$ depends on the stiffness of the strip but is
independent of the elastic properties of the spherical membrane to
lowest order, giving this geometry a potential advantage as a
metrological approach for determining the bending modulus of thin
stiff films.  We further show how defects in the wrinkling pattern
form when the pressure is reduced rapidly.  We also introduce a
numerical model for the system, which we employ to explore
hierarchical structures that can emerge with large reductions in
pressure.

\section{Calculation of the Instability Wavenumber}

In considering the energetics of the system, we approximate the
sphere to be perfectly flexible, so that no bending energy is
associated with its deformation, while assuming the strip is
inextensible so that its length is conserved.  We further
parameterize the distance from the center of the sphere to the strip
$r$ in polar coordinates, $r(\phi)$, and assume that the wrinkling
instability forms a sinusoidal deformation in the strip,
$r(\phi)=\bar{r}+a\cos{n\phi}$, where $n$ specifies the wavenumber.
The objective of the calculation is to determine the wavenumber at the initial 
instability, where $a$ grow infinitesimally from $a=0$.  In
particular, we assume $na\ll R$ for any finite $n$, where $R$ is
the initial radius of the balloon.  In order to find the wavenumber, we
need to calculate the total energy of the system to $\mathcal O(a^2)$.  
The inextensibility of the strip implies that
    \begin{eqnarray}
        2\pi R  &=& \int_0^{2\pi R} \mathrm d l =
        \int_0^{2\pi}\mathrm{d}\phi~\sqrt{r^2+ {r'}^2}
        = 2\pi\bar{r}+\frac{\pi a^2}{2\bar{r}}n^2 + \mathcal O(a^3).
    \end{eqnarray}
    where $dl = \sqrt{r^2 + {r'}^2} \, \mathrm d \phi$ is the
    element of strip length, $r' \equiv
    \mathrm{d}r\mathrm{/d}\phi$, and we make use of the infinitesimal nature of $a$ to expand the integrand.  We thus obtain
    \begin{equation}\label{radius}
        r(\phi) = R + a\cos{n\phi} -\frac{n^2 a^2}{4R} + \mathcal O(a^3).
    \end{equation}

Upon reduction of the air pressure in the balloon, the relevant
mechanical energy of the bilayer has three separable contributions.
The first is the energy associated with bending the strip, which has
the form \begin{equation}\label{e1}
            E_\mathrm{bending} =
            \frac{B}{2}\int\kappa^2\mathrm{d}l 
        \end{equation}
 where $B$ is the strip's bending modulus and $\kappa$ is the local
 curvature,
        \begin{equation}\label{dadl}
            \kappa=\frac{r^2+2r'^2-rr''}{(r^2+r'^2)^{3/2}}.
        \end{equation}
Inputting Eq.~(\ref{dadl}) in (\ref{e1}) and again expanding in powers
of $a$, we obtain
        \begin{eqnarray}
            E_\mathrm{bending} = \frac{B}{2} \int_0^{2\pi} \mathrm{d} \phi
            \frac{(r^2+2r'^2-rr'')^2}{(r^2+r'^2)^{\frac{5}{2}}}
                =  \frac{\pi B}{R} + \frac{\pi B a^2}{2R^3} (n^2-1)^2 
            + \mathcal O(a^3).
        \end{eqnarray}
We note that the first harmonic ($n=1$) has zero bending energy
because it amounts to a mere shift of the strip in space as long as $a
<< R$, and the lowest harmonic affecting the shape of the strip is
$n=2$.

The second contribution to the energy arises from the inward force
that the rubber at the edges of the strip applies as the sphere
deflates and its equilibrium radius decreases.  This force per unit
length (two-dimensional analog of pressure) is
$p=-2\sigma\sin{\alpha}$ where $\alpha$ is the angle between the
strip and the rubber membrane at the edge of the strip, as
illustrated in the inset to Fig.~2(a), and $\sigma$ is the tension
of the balloon.  The associated energy is
        \begin{eqnarray}
            E_\mathrm{pressure} =  -pA = 
            -\frac{p}{2}\int_0^{2\pi}\mathrm{d}\phi~r(\phi)^2
            = 2\pi R^2 \sigma \sin{\alpha} 
            - \pi (n^2-1) a^2 \sigma \sin{\alpha}
            + \mathcal O(a^3),
        \end{eqnarray}
where $A$ is the area enclosed by the strip.  Again, the first
harmonic costs no energy because it does not affect the shape of the
strip.  We note that this expression is strictly correct only prior to
the initial instability.  After the instability $p$ should acquire an
dependence on $\phi$ that could affect the subsequent development of
the instability but should not impact the instability wavenumber at
the onset.
        
If the total energy of the bilayer were only $E_\mathrm{bending} +
E_\mathrm{pressure}$, then the minimum energy solution for buckling
would correspond to the lowest available harmonic, $n=2$, and the
instability would follow the analog of the Euler result for a
circle: the strip would buckle into an ellipse.  However, the
elastic nature of the sphere changes this situation in a dramatic
way because the additional stretching of the membrane introduced by
a sinusoidal deformation penalizes large amplitude, long-wavelength
deformations. To evaluate this elastic contribution, we consider the
limit where the radius of the balloon is much larger than the width
of the strip, so that the region of the rubber membrane affected by
the strip  can be treated essentially as a cylindrically symmetric
system of radius $\bar{r}$.  The solution we will reach is
consistent with this approximation.  As long as the deformations of
the membrane are small compared to the radius, the shape of the
elastic membrane near the strip will obey Laplace's equation
$\nabla^2 u=0$~\cite{fulop}. In cylindrical coordinates this is
        \begin{equation}\label{Laplace}
            \frac{\mathrm{d}^2 u}{\mathrm{d}z^2}
            +\frac{1}{\bar{r}^2}\frac{\mathrm{d}^2 u}{\mathrm{d}\phi^2}
            = 0,
        \end{equation}
where $z$ specifies the direction along the balloon perpendicular to
the strip and $u$ is the local deviation of the distance from the
center to the surface of the membrane away from the average radius.
The energy associated with the elastic deformation of the rubber is
        \begin{equation}\label{e3}
        E_\mathrm{elastic}
         =\int_0^{2\pi} \bar{r}\, \mathrm{d}\phi\int\mathrm{d}z \,
        \frac{\sigma}{2}
        \left[
          \left(\frac{\mathrm{d} u}{\mathrm{d}z}\right)^2+
           \frac{1}{\bar{r}^2}
            \left(\frac{\mathrm{d} u}{\mathrm{d}\phi}\right)^2
        \right],
        \end{equation}
where $\sigma$ is again the tension of the sphere.  Using the ansatz
$u(z,\phi)=f(z)a\cos{n\phi}$, as suggested by Eq.~(\ref{radius}) (with
$f(0)=1$ and $f(\pm \infty)=0$), Laplace's equation becomes
        \begin{equation}
            \frac{\mathrm{d}^2
            f}{\mathrm{d}z^2}-\frac{n^2}{\bar{r}^2}f=0.
        \end{equation}
Hence $f(z)=\exp{(-|nz|/\bar{r})}$.  That is, the deformation of the
balloon decays exponentially as a function of distance from the paint
strip with decay length $\bar{r}/n$, and this result justifies the
cylindrical approximation as long as the instability wavenumber $n$ is
sufficiently large Using this solution in the energy of Eq.~(\ref{e3})
gives
        \begin{equation}\label{E3}
        E_\mathrm{elastic}=\pi\sigma a^2 |n| +\mathcal O(a^3).
        \end{equation}
        
Therefore, the total energy we must consider is $E(a,n) =
E_\mathrm{bending} + E_\mathrm{pressure} + E_\mathrm{elastic}$.
Anticipating that the instability occurs at a high wavenumber we
neglect the difference between $n^2-1$ and $n^2$.  In this
approximation the total energy of the strip to $\mathcal O(a^2)$ is
         \begin{equation}\label{E}
                E(a,n) = \frac{\pi a^2}{2}\left(
                  Bn^4/R^3 - 2n^2 \sigma \sin{\alpha} + 2\sigma |n|
                \right).
         \end{equation}
As a function of $n$, this energy has a minimum at $n=n_0$ given by
the condition $\partial E/\partial n = 0$.  The instability occurs
when $E(a,n_0) = 0$.  Thus, we obtain the set of equations
         \begin{eqnarray}
         Bn^4/R^3 - 2n^2 \sigma \sin{\alpha} + 2\sigma n &=& 0,\nonumber\\
         4Bn^3/R^3 - 4n \sigma \sin{\alpha} + 2\sigma  &=& 0.
         \end{eqnarray}
Solving these, we find that the instability occurs at the wavenumber
         \begin{equation}\label{n0}
         n_0 = \left(\frac{\sigma R^3}{B}\right)^{1/3}.
         \end{equation}
The balloon pulls on the strip at an angle $\alpha$ such that
         \begin{equation}
         \sin{\alpha} = \frac{3}{2n}.
         \end{equation}
The angle is small if $n \gg 1$.
         
\section{Measurements of Initial Instability}

To compare the wavenumber of the instability predicted by
Eq.~(\ref{n0}) with that produced experimentally, we fabricated a
set of bilayer membranes like those shown in Fig.~1.   Nominally
spherical balloons were inflated to an initial pressure of
approximately $3100$ Pa, corresponding to an initial radius of
approximately 0.15 m.  Paint (Krylon, ``Fusion for Plastic'') was
applied to each balloon in a strip about the equator.  The strip
width $w$ was kept sufficiently narrow, $w=5$ mm, to neglect the
curvature of the balloon across the width.  To assure uniform
application of the paint, each balloon was mounted on a turntable,
and the paint was applied with an aerosol spray can positioned 0.3 m
from the balloon, while the balloon rotated at an angular velocity
of 60 RPM.  Paper shields taped to the balloons exposed the 5 mm
wide region around the balloons' equator to the paint.  To tune the
thickness and hence the bending rigidity of the strip, the paint was
applied in multiple thin coats, with a delay of approximately 2
minutes between each coat for the paint to set.  The final thickness
of the paint layers for different balloons was varied between 25
$\mu$m and 160 $\mu$m, while the variability in thickness of any
individual strip was less than 10\%.

After the painting, the balloons were left undisturbed for 18 hours to
allow the paint to dry fully.  To investigate the wrinkling
instability, we measured the circumference of the strip to obtain $R$
and then slowly released air from each balloon until deformations in
the strip appeared like those in Fig.~1(b).  The final air pressure
$P_f$ in the balloon was then measured.  A measuring tape was
carefully fastened to the balloon parallel to the wrinkled strip, and
we determined the final balloon radius $R_f$, which was typically
about 5\% smaller than $R$.  Digital photographs of the strips were
taken, and image analysis was performed to extract the wavelength
$\lambda$ of the undulations.  Measurements of $\lambda$ were made at
several points around the circumference of each balloon.  From these
results an average $\lambda$ as a function of thickness was obtained.
The bending modulus of a paint strip can be related to its thickness
$h$ by $B = \frac{E w h^{3}}{12(1-\nu^2)}$ where $E$ is the Young's
Modulus and $\nu$ is the Poisson ratio~\cite{landau}, which through
Eq.~(\ref{n0}) leads to the prediction
\begin{equation}\label{lambda}
\lambda=2\pi R_f/n_0 = 2\pi R_f\left(B/R^3\sigma\right)^{1/3} = 2\pi
\left(\frac{Ew}{6(1-\nu^{2})}\right)^{1/3}
\left(\frac{R_f^2h^3}{R^3P_f}\right)^{1/3}
\end{equation}
where the tension is related to the final pressure by $\sigma =
R_fP_f/2$.

\begin{figure}
\includegraphics[scale=1] {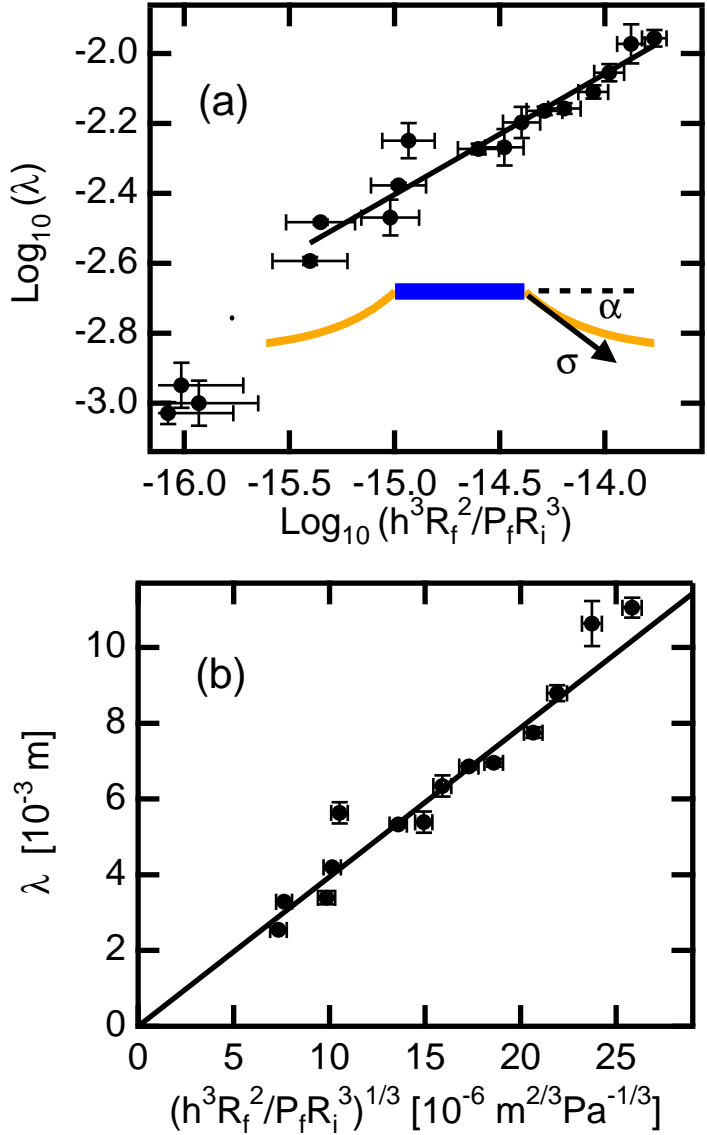}
\caption{\label{Figure 2:} (Color online) Results for the wrinkling
patterns formed in paint strips of varying thickness following balloon
deflation. In (a) the logarithm of the wavelength $\lambda$ is plotted
against the logarithm of the quantity $\frac{R_f^2h^3}{R^3P_f}$.  The
line is the result of power law fit, $\lambda \sim
\left(\frac{R_f^2h^3}{R^3P_f}\right)^{0.34 \pm 0.05}$, in good
agreement with the predicted scaling, Eq.~(\ref{lambda}).  The data at
the smallest $h$ is excluded from the fit because the inextensibility
condition is not obeyed for the thinnest paint layers.  The inset
shows a schematic cross section of the strip and adjoining rubber.
The rubber on average pulls inward on the strip with a force per unit
length $p = -2\sigma$ sin$\alpha$.  In (b) $\lambda$ is plotted
against $\left(\frac{h^3}{R_fP_f}\right)^{1/3}$ on linear scales.  The
line is the result of a linear fit from which the Young's modulus of
the paint is extracted.  } \label{fig2}
\end{figure}

Figure 2 displays the results for $\lambda$ determined from a set of
strips with different paint thicknesses.  The wavelength is plotted in
Fig.~2(a) against $R_f^2h^{3}/R^3P_f$ on logarithmic scales to test
the scaling relation predicted by Eq.~(\ref{lambda}).  The error bars
associated with the abscissa in Fig.~2 result primarily from the 10\%
variability in the paint thickness $h$, while the uncertainties in
$\lambda$ reflect the spread in the measured values for each
thickness, which again presumably results primarily from the
variability in $h$ for each balloon.  As Fig.~2(a) demonstrates, the
results for sufficiently large $h$ support the scaling prediction in
Eq~(\ref{lambda}) well, as evidenced by the result of a power-law fit,
$\lambda \sim \left(\frac{R_f^2h^3}{R^3P_f}\right)^{0.34 \pm 0.05}$,
shown by the solid line.  Excluded from the fit are three data points
at the smallest $h$, which deviate from this scaling.  Observations of
the onset of the instability indicated that the assumption of
inextensibility of the strip was not valid at these smallest values of
$h$.  Specifically, we observed that the wrinkles annealed away
moments after their formation, indicating that the thinnest paint
layers are able to contract to relieve the stress.  Relaxation of the
inextensibility condition modifies the constraint on the modulation
amplitude $a$, expressed in Eq.~(2), ultimately changing the resulting
scaling to $\lambda \sim \left(\frac{\sigma
R^3}{B}\right)^{1/2}$~\cite{paula}.  No such contraction of the paint
strip was visible at larger $h$ where the results display good
agreement with the predicted scaling.

In order to test the wavelength prediction in Eq.~(\ref{lambda})
quantitatively, we display in Fig.~2(b) $\lambda$ versus
$\left(\frac{R_f^2h^3}{R^3P_f}\right)^{1/3}$ on linear scales,
excluding the three points at smallest $h$.  Assuming a Poisson ratio
typical for an acrylic elastomer, $\nu =0.35$, we extract from the fit
the Young's modulus of the painted layer, $E= 2.7 \pm 0.3 \times 10^8$
Pa, a very reasonable value for the paint.  As a way to correlate this
value with another measurement, we performed an additional experiment
using the cantilever method to obtain the Young's modulus.  The
results of that experiment gave $E= 2.9 \pm 0.4 \times 10^8$ Pa, in
good agreement with the value extracted from the wrinkling
instability, further indicating the quantitative validity of
Eq.~(\ref{lambda}).

\begin{figure}
\includegraphics[scale=1.5] {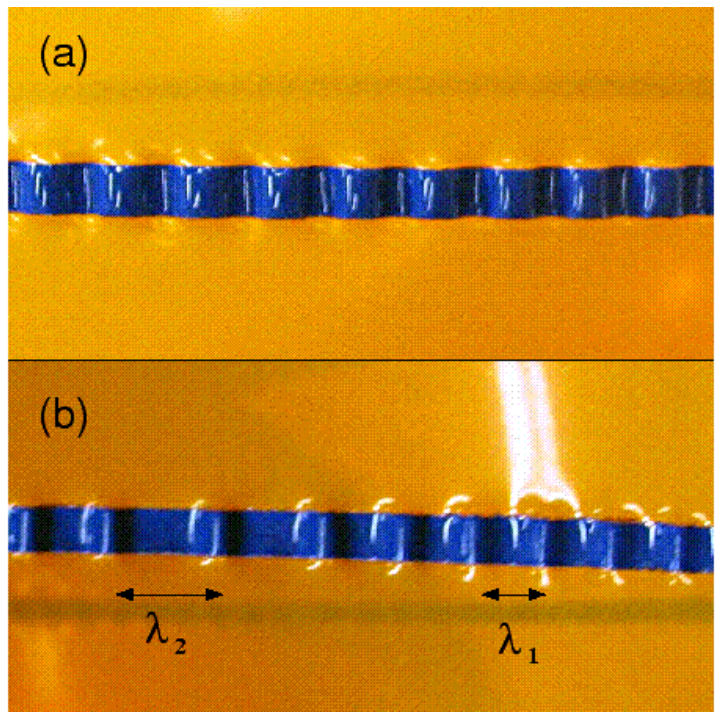}
\caption{\label{Figure 3:} (Color online) Upon rapid deflation, defects in the
wrinkling pattern appear.  (a) shows a photo of a uniform pattern
resulting from slow deflation.  The instability in (b) follows rapid
deflation, which creates segments with equilibrium wavelengths
$\lambda_1$ separated by regions of non-optimal wavelengths
$\lambda_2$.  } \label{fig3}
\end{figure}

We note that highly uniform wrinkling
patterns, like those in Fig.~1(b), form reliably when the pressure
in the balloon is decreased sufficiently slowly.  In contrast, when
the balloon is rapidly deflated, defects in the pattern appear.
Figure 3(a) shows a photograph of the highly periodic modulation
formed from slow deflation, while Fig.~3(b) shows a pattern formed
by rapidly deflating a balloon at a rate $dP/dt \approx 2000$ Pa/s.
In addition to regions with the wavelength expected from
Eq.~(\ref{lambda}), labeled $\lambda_1$, the pattern displays
regions with distinctly longer wavelength, such as $\lambda_2$.
Similar defects appear commonly in wrinkling
phenomena~\cite{bowden98,maha_natmat}.  Their appearance in the
strip following rapid deflation suggests that they are a consequence
of non-simultaneity of wrinkle nucleation in different regions
around the circumference.  When the incommensurate modulations in
such different regions meet, the adjoining segments must adopt
non-optimal wavelengths to match the regions smoothly.
Unfortunately, difficulty in controlling quantitatively the
deflation rate at high rates prevented us from investigating this
effect in detail.

\section{Second Generation Wrinkling: Experiments and Modeling}

As the pressure in the balloon is further reduced beyond the onset
of the instability, the experiments display one of two effects on
the wrinkling pattern.  For large paint thickness and hence for
large $B$,  a uniform pattern persists, and the amplitude of the
wrinkles grows until the pattern becomes highly corrugated, as shown
in Fig.~4(a).  For smaller $B$, a second larger-wavelength
modulation emerges, as shown in  Fig.~4(b).   This second wavelength
recalls the hierarchical, self-similar patterns observed recently in
the wrinkling of stiff surfaces on soft thick
substrates~\cite{maha_natmat}.

\begin{figure}
\includegraphics[scale=1.5] {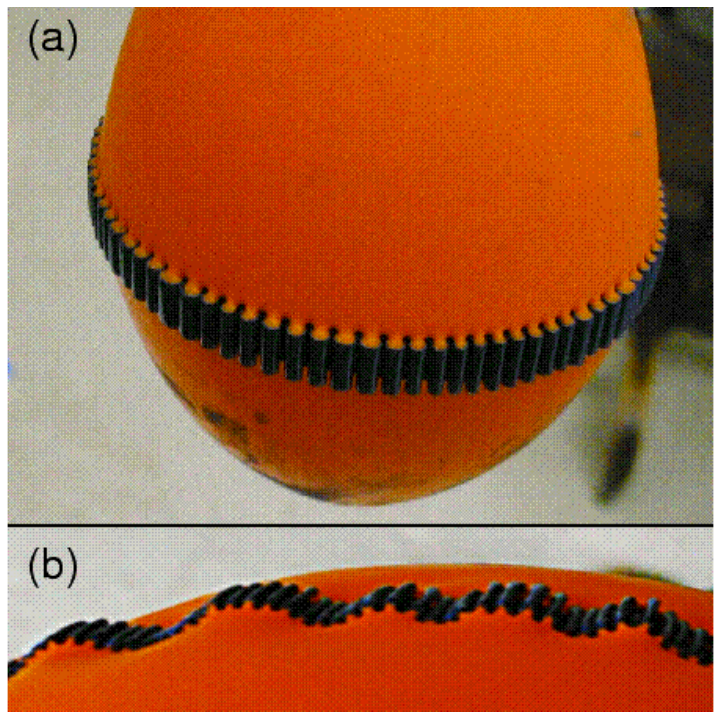}
\caption{\label{Figure 4:} (Color online) Photographs of paint strips
on balloons deflated far beyond the onset of the wrinkling
instability.  For thin layers of paint and hence small bending
modulus, the deformation remains uniform and becomes highly
corrugated, as in (a).  For thick layers and large modulus, a second,
longer wavelength instability forms, as in (b). } \label{fig4}
\end{figure}

To investigate the emergence of this
second generation of wrinkles, we simulate the bilayer with a
discrete model similar to one introduced previously to study
membrane elasticity~\cite{nelson88}. The membrane is approximated by
two concentric circular chains of particles.   Each chain has $N$
particles, with $\vec{r}_{1,i}$ and $\vec{r}_{2,i}$ specifying the
position of the $i^{th}$ particle in the outer and inner chains,
respectively.  The bending of the membrane is specified through unit
vectors $\hat{n}_{i}$ for each particle on the outer chain, where
the vector orientation is defined as normal to the line connecting
the particle's two nearest neighbors on the chain.  The membrane
energy is defined as
 \begin{equation}\label{sim}
F=\sum_{i} -J\hat{n}_{i+1}\cdot \hat{n}_{i}
+\sum_{i=1}^N\sum_{j=1}^2V_j(|\vec{r}_{j,i+1}-\vec{r}_{j,i}|-a_j)^{2}
+\sum_{i=1}^NV_3(|\vec{r}_{1,i}-\vec{r}_{2,i}|-a_3)^{2}
\end{equation}
where the first term sets the bending energy, with $J$ determining the
bending modulus, and the second term specifies the stretching energy,
with spring constants $V_1$ and $V_2$ determining the compression
modulus of the inner and outer chains, respectively. The third term
couples each particle to its neighbor in the other chain through a
spring force with constant $V_3$, and the equilibrium bond length
$a_3$ sets the thickness of the membrane. The simulations begin with
both chains forming undistorted circles, which minimizes both the
bending and stretching energies, and with the inner circle having a
radius $R_0$.  The initial equilibrium bond lengths $a_1$ and $a_2$
are set at values determined by $R_0$ and $a_3$ such that the initial
stretching energy is zero.  To simulate the shrinkage of the inner
layer, we decrease the equilibrium bond length in the inner chain in
small steps $\delta a_2$ keeping all other variables constant.
Typically, $\delta a_2/a_2^0 = 10^{-5}$, where $a_2^0$ is the initial
bond length. After each step, the particle positions, $\vec{r}_{1,i}$
and $\vec{r}_{2,i}$, are adjusted to minimize the energy,
Eq.~(\ref{sim}), using the conjugate gradient method~\cite{numrec}.
After a certain number of steps, the circular symmetry is suddenly
broken and the membrane deforms with a small amplitude, small
wavelength pattern.  This wrinkling is preferred energetically over
deformation into an Euler-like ellipse because the small wavelength
pattern enables the two chains to maintain a larger length difference,
thus reducing their respective stretching energies, while avoiding a
large separation between chains that would contribute a large energy
from the term coupling the chains together.  Thus, in the simulations
the coupling between the inner and outer chains provides the
constraint that prohibits large amplitude deformations.

We find the wrinkling of the simulated membranes displays many of the
salient features observed experimentally.  For example, defects like
those in Fig.~3(b) develop when the step size for the shrinkage
$\delta a_2/a_2^0$ is made large.  The wavelength in the simulations
is observed to scale with bending constant as $\lambda \sim J^{1/2}$,
in contrast to the scaling predicted by Eq.~(\ref{lambda}) and
observed in Fig.~2(a).  We ascribe this difference to the
extensibility of the outer chain in the simulations, which as
described above is expected to alter the scaling.  A detailed
description of these findings will be presented
elsewhere~\cite{paula}.  Here, we focus on the behavior as the
equilibrium radius of the inner chain is further reduced beyond the
onset of the instability.  In agreement with experiment, the amplitude
of the wrinkles grows, and eventually a second generation of
longer-wavelength wrinkles suddenly appears only if the bending
constant is sufficiently small.  The onsets of the first and second
generation instabilities are easily identified by the sudden
appearance of pronounced peaks in the Fourier power spectrum of the
particle displacements.  The inset to Fig.~5 shows examples of such
Fourier spectra shortly before and immediately after the formation of
second generation wrinkles in a simulation.

\begin{figure}
\includegraphics[scale=1] {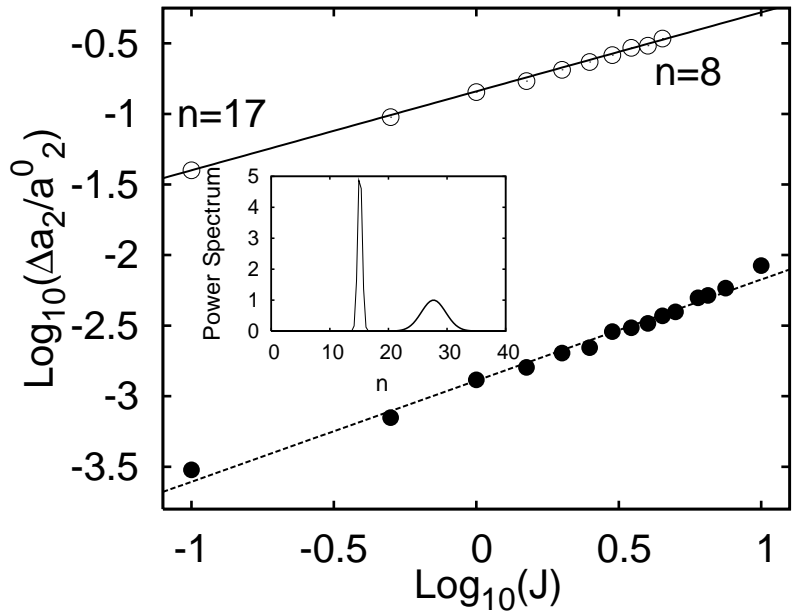}
\caption{\label{Figure 5:} Simulation results for the onset of the
wrinkling instability.  The relative strain $\Delta a_2/a_2^0$ at
which the first generation (solid) and second generation (open)
instabilities occur are displayed as a function of bending constant
$J$.  The simulation parameters employed were $N=200$, $a_3=0.3$,
$R_0=15$, and $V_j=50,2,30$ for $j=1,2,3$ respectively.   The inset
shows the Fourier power spectrum of the particle displacements
shortly before and immediately after the formation of second
generation wrinkles in a simulation using the same parameters except
$N=600$ and $J=0.5$.  The first generation power-spectrum peak ($n
\approx 27$) has been scaled by $10^7$ for clarity. } \label{fig5}
\end{figure}

Figure \ref{fig5} displays the total change in $a_2$ normalized by $a_2^0$
at which the first and second generation wrinkle patterns appear in
a  series of simulations with varying bending constant $J$.  Above a
certain value of $J$, the second generation of wrinkles never
appears, and instead the amplitude of the first generation continues
to grow until the simulations break down. (Breakdown occurs when the
inner chain passes through the outer chain, reminiscent of the
experimental situation at large $B$ in which the first generation
modulation grows to the point that $r(\phi)$ is no longer single
valued.)  Strikingly, the strain amplitudes at which the first and
second generation wrinkles appear display very similar
quasi-power-law dependence on $J$.  As suggested
previously~\cite{maha_natmat},  the membrane with first generation
wrinkles can be viewed as a membrane with a larger effective
thickness and larger effective bending modulus, which in turn
wrinkles with a larger wavelength to create the hierarchical
pattern.  The similarity between the curves in Fig.~5 lends support
to this picture, suggesting that the second generation instability
can be mapped to the first through a simple renormalization of $J$
that would lead naturally to self-similar structures.

\section{Concluding Remarks}

In summary, this simple system of a stiff strip on a spherical
membrane has allowed us to investigate in detail a buckling
instability in which a constraint enforces a small wavelength
deformation.  In this case, the constraint derived from the large
energy cost of membrane stretching that would accompany an
Euler-like, large amplitude buckling deformation.  As a final
perspective on the consequences of this constraint, we note briefly
that  one can formulate this contribution to the energy in terms of
a long-range interaction between local deformations of the different
parts of the strip. Specifically,  above we analyzed the energy of
small fluctuations of the local strip radius $u(\phi) = r(\phi) -
\bar{r}$ for a given harmonic $u(\phi) = a \cos{n\phi}$.  In a more
general case the total energy (\ref{E}) can be written in terms of
the Fourier transform of $u(l)$:
\begin{equation}
E[u(l)] = 2\pi R \int \frac{\mathrm d k}{2\pi} \left( B k^4/2 + k^2
-R \sigma \sin{\alpha} + \sigma |k|\right)|u_k|^2.
\end{equation}
where $l$ is the position measured along the strip.  The first two
terms -- the bending and pressure energies -- have regular Fourier
transforms, whereas the elastic term has a kink singularity at
$k=0$.  Returning to real space we find that the bending and
pressure energies become local terms containing spatial derivatives.
The singular elastic term translates into a long-range interaction:
\begin{equation}
E[u(l)] = \int \mathrm d l \left[ \frac{B}{2} \left( \frac{\mathrm
d^2 u}{dl^2} \right)^2
 - R \sigma \sin{\alpha} \left( \frac{\mathrm d u}{dl} \right)^2 \right]
 + \sigma \int dl_1 \int  dl_2 \
  u(l_1) V(l_1 - l_2) u(l_2),
\end{equation}
where the interaction kernel is
\begin{equation}
V(x) =   \frac{\delta(x)}{t} - \frac{1}{\pi(x^2 + t^2)},
\end{equation}
and the parameter $t$ is an ultraviolet cutoff regularizing the
short-distance behavior of the interaction.   This identification of
the constraint with an effective long-range interaction suggests a
method for mapping such wrinkling phenomena onto other condensed
matter systems where competition between long-range and short-range
interactions lead to pattern formation, such as for example in the
creation of stripe domains in thin ferromagnetic films~\cite{ferro}.
An interesting direction for further study would be to investigate
whether such an approach can aid in understanding and predicting
wrinkling behavior in less simple situations and further whether it
might lend insight into cases in nature, such as the buckling of
biological membranes during growth~\cite{drasdo2000}, in which the
membranes are observed to wrinkle. 

\vspace{1 cm} 
\noindent 
{\bf Acknowledgments: } 
We thank A. Belmonte and R. Cammarata for helpful
discussions and L. Sung for useful correspondence.  Funding was
provided by the NSF under grants DMR-0134377 and DMR-0520491.

\newpage


\begin{thebibliography}{19}
\expandafter\ifx\csname natexlab\endcsname\relax\def\natexlab#1{#1}\fi
\expandafter\ifx\csname bibnamefont\endcsname\relax
  \def\bibnamefont#1{#1}\fi
\expandafter\ifx\csname bibfnamefont\endcsname\relax
  \def\bibfnamefont#1{#1}\fi
\expandafter\ifx\csname citenamefont\endcsname\relax
  \def\citenamefont#1{#1}\fi
\expandafter\ifx\csname url\endcsname\relax
  \def\url#1{\texttt{#1}}\fi
\expandafter\ifx\csname urlprefix\endcsname\relax\def\urlprefix{URL }\fi
\providecommand{\bibinfo}[2]{#2}
\providecommand{\eprint}[2][]{\url{#2}}

\bibitem[{\citenamefont{Gladden et~al.}(2005)\citenamefont{Gladden, Handzy,
  Belmonte, and Villermaux}}]{gladden}
\bibinfo{author}{\bibfnamefont{J.~R.} \bibnamefont{Gladden}},
  \bibinfo{author}{\bibfnamefont{N.~Z.} \bibnamefont{Handzy}},
  \bibinfo{author}{\bibfnamefont{A.}~\bibnamefont{Belmonte}}, \bibnamefont{and}
  \bibinfo{author}{\bibfnamefont{E.}~\bibnamefont{Villermaux}},
  \bibinfo{journal}{Phys. Rev. Lett.} \textbf{\bibinfo{volume}{94}},
  \bibinfo{pages}{035503} (\bibinfo{year}{2005}).

\bibitem[{\citenamefont{Brangwynne et~al.}(2006)\citenamefont{Brangwynne,
  MacKintosh, Kumar, Geisse, Talbot, Mahadevan, Parker, Ingber, and
  Weitz}}]{brangwynne}
\bibinfo{author}{\bibfnamefont{C.~P.} \bibnamefont{Brangwynne}},
  \bibinfo{author}{\bibfnamefont{F.~C.} \bibnamefont{MacKintosh}},
  \bibinfo{author}{\bibfnamefont{S.}~\bibnamefont{Kumar}},
  \bibinfo{author}{\bibfnamefont{N.~A.} \bibnamefont{Geisse}},
  \bibinfo{author}{\bibfnamefont{J.}~\bibnamefont{Talbot}},
  \bibinfo{author}{\bibfnamefont{L.}~\bibnamefont{Mahadevan}},
  \bibinfo{author}{\bibfnamefont{K.~K.} \bibnamefont{Parker}},
  \bibinfo{author}{\bibfnamefont{D.~E.} \bibnamefont{Ingber}},
  \bibnamefont{and} \bibinfo{author}{\bibfnamefont{D.~A.} \bibnamefont{Weitz}},
  \bibinfo{journal}{J. Cell Biol.} \textbf{\bibinfo{volume}{173}},
  \bibinfo{pages}{733} (\bibinfo{year}{2006}).

\bibitem[{\citenamefont{Allen}(1969)}]{allenbook}
\bibinfo{author}{\bibfnamefont{H.~G.} \bibnamefont{Allen}},
  \emph{\bibinfo{title}{Analysis and Design of Structural Sandwich Panels}}
  (\bibinfo{publisher}{Pergamon}, \bibinfo{address}{New York},
  \bibinfo{year}{1969}).

\bibitem[{\citenamefont{Bowden et~al.}(1998)\citenamefont{Bowden, Brittain,
  Evans, Hutchinson, and Whitesides}}]{bowden98}
\bibinfo{author}{\bibfnamefont{N.}~\bibnamefont{Bowden}},
  \bibinfo{author}{\bibfnamefont{S.}~\bibnamefont{Brittain}},
  \bibinfo{author}{\bibfnamefont{A.~G.} \bibnamefont{Evans}},
  \bibinfo{author}{\bibfnamefont{J.~W.} \bibnamefont{Hutchinson}},
  \bibnamefont{and} \bibinfo{author}{\bibfnamefont{G.~M.}
  \bibnamefont{Whitesides}}, \bibinfo{journal}{Nature}
  \textbf{\bibinfo{volume}{393}}, \bibinfo{pages}{146} (\bibinfo{year}{1998}).

\bibitem[{\citenamefont{Groenewold}(2001)}]{jan2001}
\bibinfo{author}{\bibfnamefont{J.}~\bibnamefont{Groenewold}},
  \bibinfo{journal}{Physica A} \textbf{\bibinfo{volume}{298}},
  \bibinfo{pages}{32} (\bibinfo{year}{2001}).

\bibitem[{\citenamefont{Cerda et~al.}(2002)\citenamefont{Cerda, Ravi-Chandar,
  and Mahadevan}}]{cerda02}
\bibinfo{author}{\bibfnamefont{E.}~\bibnamefont{Cerda}},
  \bibinfo{author}{\bibfnamefont{K.}~\bibnamefont{Ravi-Chandar}},
  \bibnamefont{and}
  \bibinfo{author}{\bibfnamefont{L.}~\bibnamefont{Mahadevan}},
  \bibinfo{journal}{Nature} \textbf{\bibinfo{volume}{419}},
  \bibinfo{pages}{579} (\bibinfo{year}{2002}).

\bibitem[{\citenamefont{Cerda and Mahadevan}(2003)}]{mahawrinkle}
\bibinfo{author}{\bibfnamefont{E.}~\bibnamefont{Cerda}} \bibnamefont{and}
  \bibinfo{author}{\bibfnamefont{L.}~\bibnamefont{Mahadevan}},
  \bibinfo{journal}{Phys. Rev. Lett.} \textbf{\bibinfo{volume}{90}},
  \bibinfo{pages}{074302} (\bibinfo{year}{2003}).

\bibitem[{\citenamefont{Efimenko et~al.}(2005)\citenamefont{Efimenko,
  Rackaitis, Manias, Vaziri, Mahadevan, and Genzer}}]{maha_natmat}
\bibinfo{author}{\bibfnamefont{K.}~\bibnamefont{Efimenko}},
  \bibinfo{author}{\bibfnamefont{M.}~\bibnamefont{Rackaitis}},
  \bibinfo{author}{\bibfnamefont{E.}~\bibnamefont{Manias}},
  \bibinfo{author}{\bibfnamefont{A.}~\bibnamefont{Vaziri}},
  \bibinfo{author}{\bibfnamefont{L.}~\bibnamefont{Mahadevan}},
  \bibnamefont{and} \bibinfo{author}{\bibfnamefont{J.}~\bibnamefont{Genzer}},
  \bibinfo{journal}{Nature Mater.} \textbf{\bibinfo{volume}{4}},
  \bibinfo{pages}{293} (\bibinfo{year}{2005}).

\bibitem[{\citenamefont{Harrison et~al.}(2004)\citenamefont{Harrison, Stafford,
  Zhang, and Karim}}]{harrison}
\bibinfo{author}{\bibfnamefont{C.}~\bibnamefont{Harrison}},
  \bibinfo{author}{\bibfnamefont{C.~M.} \bibnamefont{Stafford}},
  \bibinfo{author}{\bibfnamefont{W.}~\bibnamefont{Zhang}}, \bibnamefont{and}
  \bibinfo{author}{\bibfnamefont{A.}~\bibnamefont{Karim}},
  \bibinfo{journal}{Appl. Phys. Lett.} \textbf{\bibinfo{volume}{85}},
  \bibinfo{pages}{4016} (\bibinfo{year}{2004}).

\bibitem[{\citenamefont{Lacour et~al.}(2003)\citenamefont{Lacour, Wagner,
  Huang, and Suo}}]{lacour}
\bibinfo{author}{\bibfnamefont{S.~P.} \bibnamefont{Lacour}},
  \bibinfo{author}{\bibfnamefont{S.}~\bibnamefont{Wagner}},
  \bibinfo{author}{\bibfnamefont{Z.}~\bibnamefont{Huang}}, \bibnamefont{and}
  \bibinfo{author}{\bibfnamefont{Z.}~\bibnamefont{Suo}},
  \bibinfo{journal}{Appl. Phys. Lett.} \textbf{\bibinfo{volume}{82}},
  \bibinfo{pages}{2404} (\bibinfo{year}{2003}).

\bibitem[{\citenamefont{Khang et~al.}(2006)\citenamefont{Khang, Jiang, Huang,
  and Rogers}}]{khang}
\bibinfo{author}{\bibfnamefont{D.-Y.} \bibnamefont{Khang}},
  \bibinfo{author}{\bibfnamefont{H.}~\bibnamefont{Jiang}},
  \bibinfo{author}{\bibfnamefont{Y.}~\bibnamefont{Huang}}, \bibnamefont{and}
  \bibinfo{author}{\bibfnamefont{J.~A.} \bibnamefont{Rogers}},
  \bibinfo{journal}{Science} \textbf{\bibinfo{volume}{311}},
  \bibinfo{pages}{208} (\bibinfo{year}{2006}).

\bibitem[{\citenamefont{Stafford et~al.}(2004)\citenamefont{Stafford, Harrison,
  Beers, Karim, Amis, Vanlandingham, Kim, Volksen, Miller, and
  Simonyi}}]{metrology}
\bibinfo{author}{\bibfnamefont{C.~M.} \bibnamefont{Stafford}},
  \bibinfo{author}{\bibfnamefont{C.}~\bibnamefont{Harrison}},
  \bibinfo{author}{\bibfnamefont{K.~L.} \bibnamefont{Beers}},
  \bibinfo{author}{\bibfnamefont{A.}~\bibnamefont{Karim}},
  \bibinfo{author}{\bibfnamefont{E.~J.} \bibnamefont{Amis}},
  \bibinfo{author}{\bibfnamefont{M.~R.} \bibnamefont{Vanlandingham}},
  \bibinfo{author}{\bibfnamefont{H.}~\bibnamefont{Kim}},
  \bibinfo{author}{\bibfnamefont{W.}~\bibnamefont{Volksen}},
  \bibinfo{author}{\bibfnamefont{R.~D.} \bibnamefont{Miller}},
  \bibnamefont{and} \bibinfo{author}{\bibfnamefont{E.~E.}
  \bibnamefont{Simonyi}}, \bibinfo{journal}{Nature Mater.}
  \textbf{\bibinfo{volume}{3}}, \bibinfo{pages}{545} (\bibinfo{year}{2004}).

\bibitem[{\citenamefont{Fulop}(1955)}]{fulop}
\bibinfo{author}{\bibfnamefont{W.}~\bibnamefont{Fulop}}, \bibinfo{journal}{Br.
  J. Appl. Phys.} \textbf{\bibinfo{volume}{6}}, \bibinfo{pages}{21}
  (\bibinfo{year}{1955}).

\bibitem[{\citenamefont{Landau and Lifshitz}(1986)}]{landau}
\bibinfo{author}{\bibfnamefont{L.~D.} \bibnamefont{Landau}} \bibnamefont{and}
  \bibinfo{author}{\bibfnamefont{E.~M.} \bibnamefont{Lifshitz}},
  \emph{\bibinfo{title}{Theory of Elasticity}} (\bibinfo{publisher}{Pergamon},
  \bibinfo{address}{New York}, \bibinfo{year}{1986}).

\bibitem[{\citenamefont{Concha et~al.}()\citenamefont{Concha, Mellado, McIver,
  and Leheny}}]{paula}
\bibinfo{author}{\bibfnamefont{A.}~\bibnamefont{Concha}},
  \bibinfo{author}{\bibfnamefont{P.}~\bibnamefont{Mellado}},
  \bibinfo{author}{\bibfnamefont{J.~W.} \bibnamefont{McIver}},
  \bibnamefont{and} \bibinfo{author}{\bibfnamefont{R.~L.}
  \bibnamefont{Leheny}}, \eprint{in preparation}.

\bibitem[{\citenamefont{Seung and Nelson}(1988)}]{nelson88}
\bibinfo{author}{\bibfnamefont{H.~S.} \bibnamefont{Seung}} \bibnamefont{and}
  \bibinfo{author}{\bibfnamefont{D.~R.} \bibnamefont{Nelson}},
  \bibinfo{journal}{Phys. Rev. A} \textbf{\bibinfo{volume}{38}},
  \bibinfo{pages}{1005} (\bibinfo{year}{1988}).

\bibitem[{\citenamefont{Press et~al.}(1986)\citenamefont{Press, Flannery,
  Teukolsky, and Vetterling}}]{numrec}
\bibinfo{author}{\bibfnamefont{W.~H.} \bibnamefont{Press}},
  \bibinfo{author}{\bibfnamefont{B.~P.} \bibnamefont{Flannery}},
  \bibinfo{author}{\bibfnamefont{S.~A.} \bibnamefont{Teukolsky}},
  \bibnamefont{and} \bibinfo{author}{\bibfnamefont{W.~T.}
  \bibnamefont{Vetterling}}, \emph{\bibinfo{title}{Numerical recipes: the art
  of scientific computing}} (\bibinfo{publisher}{Cambridge},
  \bibinfo{address}{New York}, \bibinfo{year}{1986}).

\bibitem[{\citenamefont{Kashuba and Pokrovsky}(1993)}]{ferro}
\bibinfo{author}{\bibfnamefont{A.~B.} \bibnamefont{Kashuba}} \bibnamefont{and}
  \bibinfo{author}{\bibfnamefont{V.~L.} \bibnamefont{Pokrovsky}},
  \bibinfo{journal}{Phys. Rev. B} \textbf{\bibinfo{volume}{48}},
  \bibinfo{pages}{10335} (\bibinfo{year}{1993}).

\bibitem[{\citenamefont{Drasdo}(2000)}]{drasdo2000}
\bibinfo{author}{\bibfnamefont{D.}~\bibnamefont{Drasdo}},
  \bibinfo{journal}{Phys. Rev. Lett.} \textbf{\bibinfo{volume}{84}},
  \bibinfo{pages}{4244} (\bibinfo{year}{2000}).

\end{thebibliography}
\end{document}